\documentclass[fleqn,10pt]{wlscirep}

\usepackage[version=4]{mhchem}
\usepackage[utf8]{inputenc}

\usepackage{amsmath,amssymb,amsfonts}%
\usepackage{amsthm}
\usepackage{mathrsfs}
\usepackage{empheq}

\usepackage[sc]{mathpazo}

\usepackage[super]{nth}

\usepackage{bm}

\usepackage{color}

\usepackage{extarrows}

\usepackage{verbatim}

\usepackage{lineno}

\usepackage{siunitx}

\usepackage{stfloats}

\usepackage{bm}

\usepackage{graphicx}

\usepackage{caption}
\usepackage{subcaption}

\usepackage{multicol}

\usepackage{multirow}

\usepackage{mhchem}

\usepackage[title]{appendix}

\usepackage{bm}

\usepackage{float}

\usepackage{threeparttable}

\renewcommand{\eqref}[1]{Eq.~$($\ref{#1}$)$}

\title{ML-extendable framework for multiphysics-multiscale simulation workflow and data management using Kadi4Mat}
\author[1]{Somnath Bharech}
\author[1,*]{Yangyiwei Yang}
\author[2]{Michael Selzer}
\author[2]{Britta Nestler}
\author[1,*]{Bai-Xiang Xu}
\affil[1]{Division Mechanics of Functional Materials, Institute of Materials Science, Technical University Darmstadt, Otto-Berndt-Strasse 3, Darmstadt, 64287, Germany}
\affil[2]{Institute for Applied Materials (IAM), Karlsruhe Institute of Technology (KIT), Kaiserstrasse 12, Karlsruhe, 76131, Germany}

\affil[*]{Corresponding author(s): Y.Y. (yangyiwei.yang@mfm.tu-darmstadt.de) and B-X.X. (xu@mfm.tu-darmstadt.de)}

\begin{abstract} 
As material modeling and simulation has become vital for modern materials science, research data with distinctive physical principles and extensive volume are generally required for full elucidation of the material behavior across all relevant scales. 
Effective workflow and data management, with corresponding metadata descriptions, helps leverage the full potential of data-driven analyses for computer-aided material design.
In this work, we propose a research workflow and data management (RWDM) framework to manage complex workflows and resulting research (meta)data, while following FAIR principles.
Multiphysics-multiscale simulations for additive manufacturing investigations are treated as showcase and implemented on Kadi4Mat -- an open source research data infrastructure. The input and output data of the simulations, together with the associated setups and scripts realizing the simulation workflow, are curated in corresponding standardized Kadi4Mat records with extendibility for further research and data-driven analyses. These records are interlinked to indicate information flow and form an ontology-based knowledge graph. 
Automation scheme for performing high-throughput simulation and post-processing integrated with the proposed RWDM framework is also presented. 
\end{abstract}
\begin{document}

\flushbottom
\maketitle

\thispagestyle{empty}

\section*{Introduction}

Materials science stands at the forefront of numerous technological innovations spanning across various industries, with a particular emphasis on its engineering background. It has evolved from its empirical and experimental roots, which focused on engineering the chemical composition and the microstructure of materials to achieve specific properties tailored for certain applications, to embracing modeling and simulation as another aspect in the new century, revolutionizing the field with computer-aided material design. This modern approach significantly accelerates the lifecycle of material innovation while reducing costs, time, resources, and energy waste, marking a significant advancement in the pursuit of sustainable and smart material development \cite{gu2021material, deagen2022,frazier2014metal}. 
The vast disparity in scales and the interdisciplinary nature of material modeling and simulation present fresh challenges in this domain, as materials exhibit behaviors across a wide range of spatial and temporal scales, which collectively influence their overall properties. Addressing these phenomena demands a variety of theoretical methodologies, each adhering to certain physical principles at corresponding scale. In other words, the multiphysics and multiscale frameworks are required to fully elucidate material behaviors across all relevant scales. 
As a result, extensive data bonded with corresponding physical principles at varying scales are normally anticipated in a material modeling and simulation attempt. These data can be roughly classified into three types: (1) input data, which are the pre-requisite quantities and geometries describing the raw/pure materials, intrinsic structures and physical conditions to initiate certain physical processes at corresponding scale; (2) output data, which are the direct/post-processed quantities and geometries presenting the response of the physical process according to certain input data; (3) auxiliary/associated data, which are not related to the input/output of a simulation, but give the necessary information to reproduce the output from a given input. 
To follow the state-of-the-art FAIR (findable, accessible, interoperable and reusable/reproducible) principles for data sharing \cite{wilkinson2016fair}, all three types of data should be collected and recorded as close to the data producing source as possible \cite{garabedian2022generating}, leading to the proper design of the data infrastructure with considerations such as the efficiency, readability, extendibility, and reliability.

Meanwhile, with the rapid development of the high-performance computing clusters, it is possible to perform material simulations in a high-throughput fashion, i.e., numerous extensive simulation tasks are simultaneously executed, targeting on the objectives requesting vast volume of data, e.g., delivering the process-microstructure-property (PMP) relationships for the manufacturing of a certain material system \cite{mcelfresh2024}. For instance, in additive manufacturing realized by powder bed fusion (PBF) techniques, over one hundred process parameters directly influence the final products \cite{Oliveira2020, king2015laser}. The most critical ones, including beam power, scan speed, beam diameter, layer thickness, hatch distance, and scanning strategies, need to be adjusted for each individual build, considering the specific material and geometry. In such cases, data-driven analyses based on statistics and/or machine learning (ML) are generally adopted to extract the PMP relationships of the targeted material system. It has been proven that data management following the FAIR principles is a key to perform scalable ML-based researches, as it readily compacts data describing the raw/pure material, the process parameters and conditions, and the response/effective properties of the processed materials, achieving the data-centric ML analyses \cite{miranda2021datacentric, La2023ComparingMA}. 
Meanwhile, many modeling and simulation methods may have to be integrated as one workflow recapitulating essential factors from various scales in a single material process, it is then essential to manage not just the data involved in a simulation workflow, but also scripts or protocols that realize the workflow in an automatic way enabling the high-throughput computations (HTC) \cite{yang2024high}. This can help to adapt the established simulation workflow for similar material systems while retaining complete reproducibility, fostering collaborative research and efficient knowledge transfer.  
Beyond these, the effective management and curation of data, coupled with simulation workflows adhering to FAIR principles, is also foundational to both scientific accountability and the robust validation and verification of research findings \cite{russell2021fair}.

\begin{figure}[ht]
    \centering
    \includegraphics[width=\linewidth]{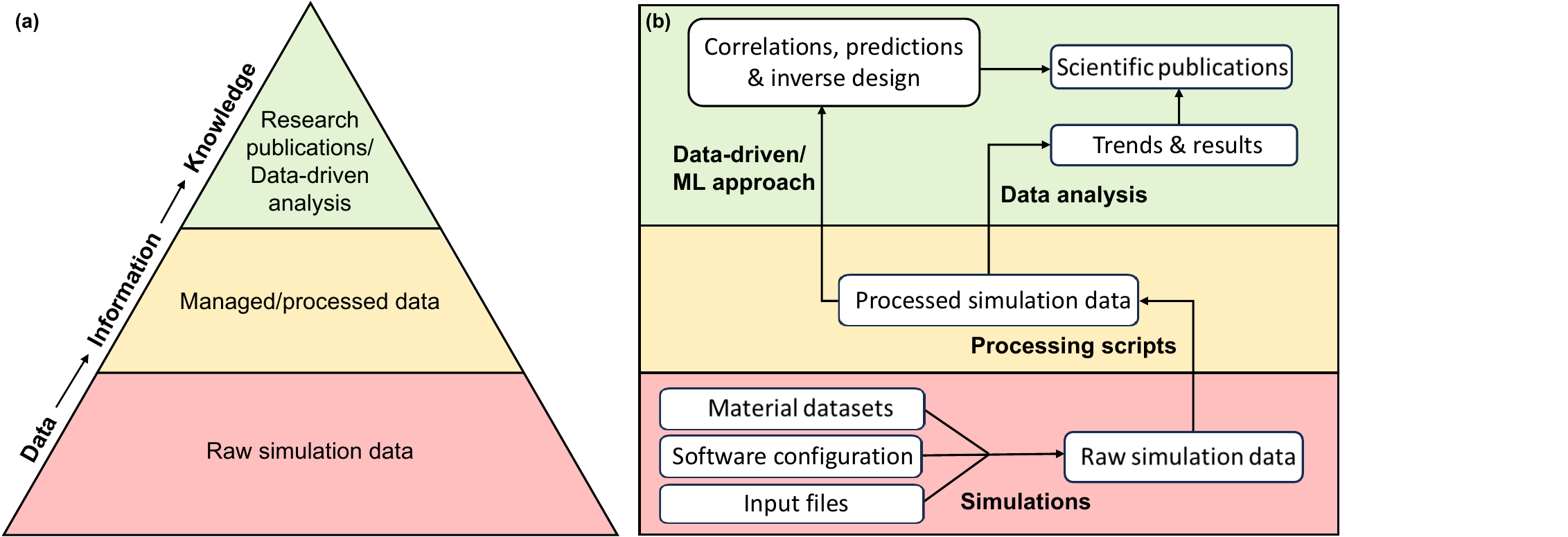}
    \caption{(a) Data-Information-Knowledge (DIK) hierarchy of research data inspired by Chaffey and Wood, 2005 (Chapter 5)\cite{chaffey2005business,rowley2007wisdom}. This hierarchy is analogous to a typical research data lifecycle, which starts as sets of discrete data. The data is processed into information and then further condensed into knowledge, which is generally documented in form of research articles. (b) Workflow of a simulation-based research, where various (meta)data aid the generation of raw simulation data. This data is further processed using scripts and macros on various software. The processed data is further analysed to form meaningful correlations, predictions and trends, which are ultimately published as scientific research.}
    \label{fig:DIK-heirarchy}
\end{figure}

Following the generalized data-information-knowledge (DIK) hierarchy as introduced by Chaffey and Wood\cite{chaffey2005business}, a comparison can be made with simulation-based investigations. As shown in the DIK hierarchy in Fig. \ref{fig:DIK-heirarchy}, data is considered to be a discrete set of facts which, when processed, transforms into information. Further analysis of this information leads to knowledge. 
This vertical transformation of data is represented using a pyramid which also signifies the condensation of volume as the data gradually transform into knowledge.
Likewise, in a typical simulation-based research, raw simulation data forms the foundation of this hierarchy and needs processing for visualization. Further analysis leads to insightful trends which are usually well-documented in form of scientific publications, as schematically represented in Fig. \ref{fig:DIK-heirarchy}. In order to maintain the comprehensiveness of the recorded knowledge, it is important to identify and recognize supporting items such as material datasets, software configuration and input parameters used in the simulations, along with pre- and post-processing scripts applied to the raw and the processed data.
This strategy for research workflow and data management (RWDM) concurs with the input-process-output (IPO) concept introduced by Griem et al\cite{griem2022kadistudio}. They describe an atomistic approach where research processes can be iteratively structured as tasks and those tasks are further represented as horizontal transformation with three generic components: (1) Input, (2) Process and (3) Output. 
Applying the IPO concept throughout the different stages of the DIK hierarchy ultimately enables us to represent the complete research process. 
Therefore, it becomes evident that an effective RWDM framework must include both horizontal as well as vertical components of a research investigation.

Kadi4Mat, the Karlsruhe Data Infrastructure for Materials Science, is an open-source research data infrastructure developed by Karlsruhe Institute of Technology\cite{brandt2021kadi4mat, kadi4mat-zenodo}. It utilizes \textit{records}, which are essentially digital objects, as the fundamental building blocks for the infrastructure to store and manage research data. The records are uniquely identified with their \textit{persistent identifiers (PID)} and can hold associated metadata alongside the data itself. Kadi4Mat offers various features to organize and manage (meta)data effectively. Records from an investigation can be grouped together to form \textit{collections} with further sub-categorization using \textit{child collections}. Additionally, customizable \textit{templates} help maintain consistency and standardization within records. The curated research data, in the form of records, can be visualized as a knowledge graph, where individual records are linked based on their relationships. This promotes data exploration and understanding of intricate relationships within the research data.
Kadi4Mat implements a role-based access control for the records. Owners can set permissions for \textit{users} or \textit{groups} based on their predefined roles like administrator, editor, collaborator and member. This selective access control ensures data security during the course of the investigation, meanwhile enhancing collaboration among researchers and scientific staff at different access levels.
On the other hand, the research data records can be published for broader accessibility, and can even be published on open repositories such as Zenodo for universal access. Apart from Kadi4Mat, there is a variety of  electronic laboratory notebook (ELN) based research data management (RDM) tools available, such as LabArchives\cite{dunie2017importance}, labfolder\cite{colabroy2019lab}, NOMAD\cite{shabih2022development} and eLabFTW\cite{carp2017elabftw, hewera2021elabftw}. Although most of them provide common functionalities for RDM as Kadi4Mat such as data integrity, data and research security, version control and team collaboration, they lack in one way or the other in comparison with Kadi4Mat. Some of them are commercial software, while Kadi4Mat is open source. In addition to serving as a ELN, Kadi4Mat also functions as the data repository distinguishing it from the RDM tools that primarily function as lab notebooks. 
In addition to the web interface, Kadi4Mat also provides programmatic access through its python-based application programming interface (API) called Kadi$^\text{APY}$\cite{kadi-apy}. This enables potentially automated interaction with Kadi4Mat using personal access tokens (PAT), facilitating seamless integration of RWDM workflows with HTC workflows. The research data generated from such investigations often needs to be exported for ML analyses. 
The data can be fetched and processed directly into the ML models/algorithms using the API access or can simply be transferred to Kadi4Mat-hosted ML utilities and applications such as KadiStudio\cite{griem2022kadistudio}, KadiAI and CIDS\cite{tosato2023bayesian}.
These features make Kadi4Mat the most appropriate infrastructure for the data management needs of complex investigations like this one.
In this work, we present a RWDM framework which is implemented for our recent numerical investigation on establishing PMP relationships during PBF process using Kadi4Mat\cite{zhou20213d, yang2023elastoplastic}. During the RWDM process of this investigation, the crucial steps involved in the workflow, the identification, collection and organization of (meta)data, their recording and crosslinking to indicate the information flow, will be discussed. Ontology-based knowledge representation of the overall investigation using the records and capturing their relation promotes further expansion and usage of the research database, as the relations are machine-readable and can be machine-actionable as well. Automation in the implementation of the proposed RWDM framework and the further usage of the curated data will also be discussed. 

\section*{Results}
\subsection*{RWDM infrastructure design}

\begin{figure}[ht]
\centering
\includegraphics[width=\linewidth]{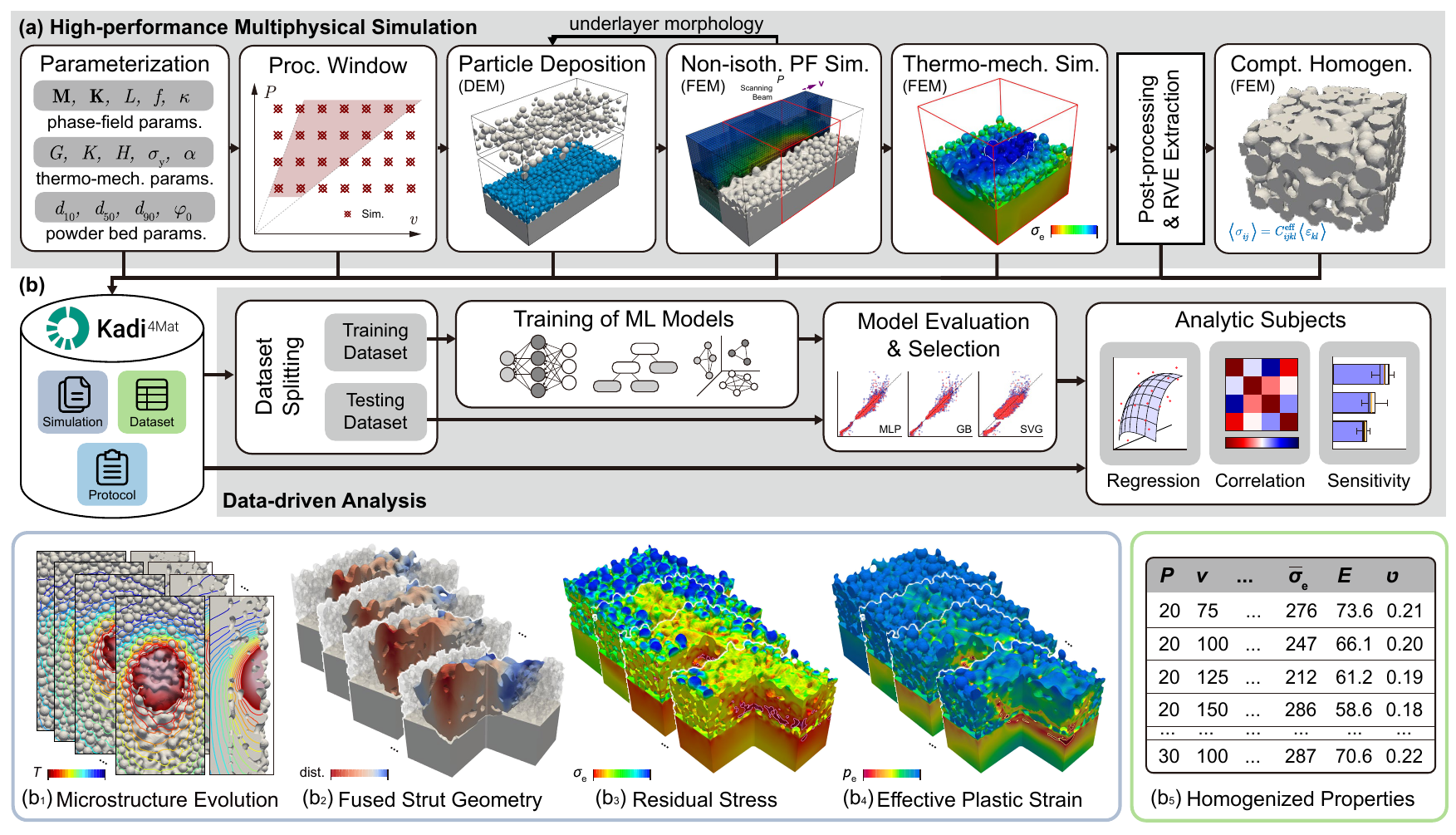}
\caption{Schematic of the research framework consisting of (a) Multiphysical simulation scheme for multilayer powder bed fusion simulations including paramterization for phase-field model, thermomechanical analysis, process parameter selection for PBF, powder bed parameters and deposition, process simulations using non-isothermal phase-field model, thermomechanical analysis, RVE selection for homogenization of mechanical properties; (b) Storing the relevant (meta)data in form of several record types, namely simulation, dataset and protocol in Kadi4Mat, to be further used in data-driven analysis.  Illustration of (b$_{1-4}$) the simulation results stored in  simulation records and (b${_5}$) data stored in form of dataset records. }
\label{fig:scheme}
\end{figure}

The RWDM framework outlined in this study comprises of workflow (including simulation  sub-routines) management, (meta)data identification and curation.
The practical implementation of this framework is illustrated by curating the research workflow and the data generated from our recent works on multilayer PBF simulations\cite{zhou20213d, yang2023elastoplastic}. 
The simulation workflow is explained in the methods section and is also visually summarized in Fig. \ref{fig:scheme}(a). Sample results from this workflow are schematically displayed in Fig. \ref{fig:scheme}(b$_{1-5}$). The four-layer PBF process simulations using non-isothermal phase-field model results in the evolution of the thermal microstructure containing the temporal information of the fused strut. Subsequent thermo-elasto-plastic calculations are performed to estimate the evolution of residual stress and plastic strain in the thermal microstructure from the process simulations.
The effective mechanical properties of the PBF processed microstructure are calculated using a computational homogenization scheme. As implied in Fig. \ref{fig:scheme}(b), the research data generated during the multilayer PBF investigation was carefully collected, organised and stored into 
Kadi4Mat in form of records. 
Three distinct record types are utilized to curate the research data from the multilayer PBF work: (1) dataset, (2) protocol and (3) simulation records. From technical point of view, these various types of records are on the same level in the RWDM infrastructure, however from managerial aspect, they are distinguished by their data content, as outlined in table \ref{tab:record-types}. The usage and integration of these different record types in this RWDM infrastructure will be further discussed in detail in the following subsection.
 
An overview of the curated data from the multilayer PBF work can be seen in form of an ontology-based knowledge graph, in Fig. \ref{fig:knowledge-graph}. The research workflow and data curated from multilayer PBF simulations consists of several records, indicated by circular nodes. These nodes can be seen arranged prominently, to visualize the child collections representing the three clusters of simulations: (1) phase-field simulations, (2) thermo-mechanical simulations and (3) computational homogenization. The records are labeled with their respective identifiers and the record types. The record type can also be identified by the node colors. 

\subsection*{Data records design}

\begin{figure}[ht]
\centering
\includegraphics[width=\linewidth]{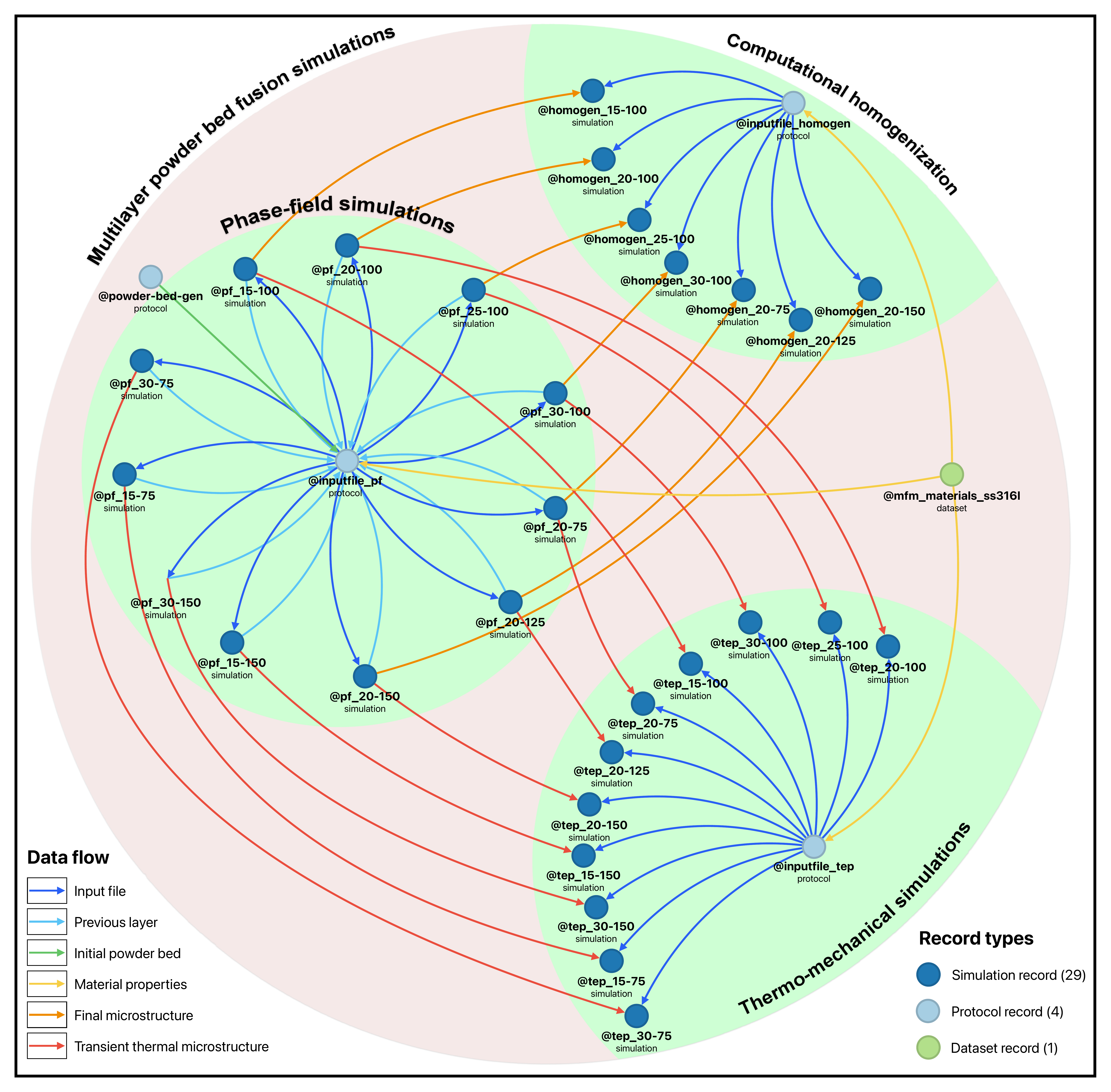}
\caption{Knowledge graph visualizing research data from multilayer PBF simulations, including phase-field simulations, computational homogenization, and thermo-mechanical analysis. The data is stored in various record types, such as simulation records, protocol records, and dataset records. The data flow within the records is also indicated.}
\label{fig:knowledge-graph}
\end{figure}

\begin{figure}
    \centering
    \includegraphics[width=\linewidth]{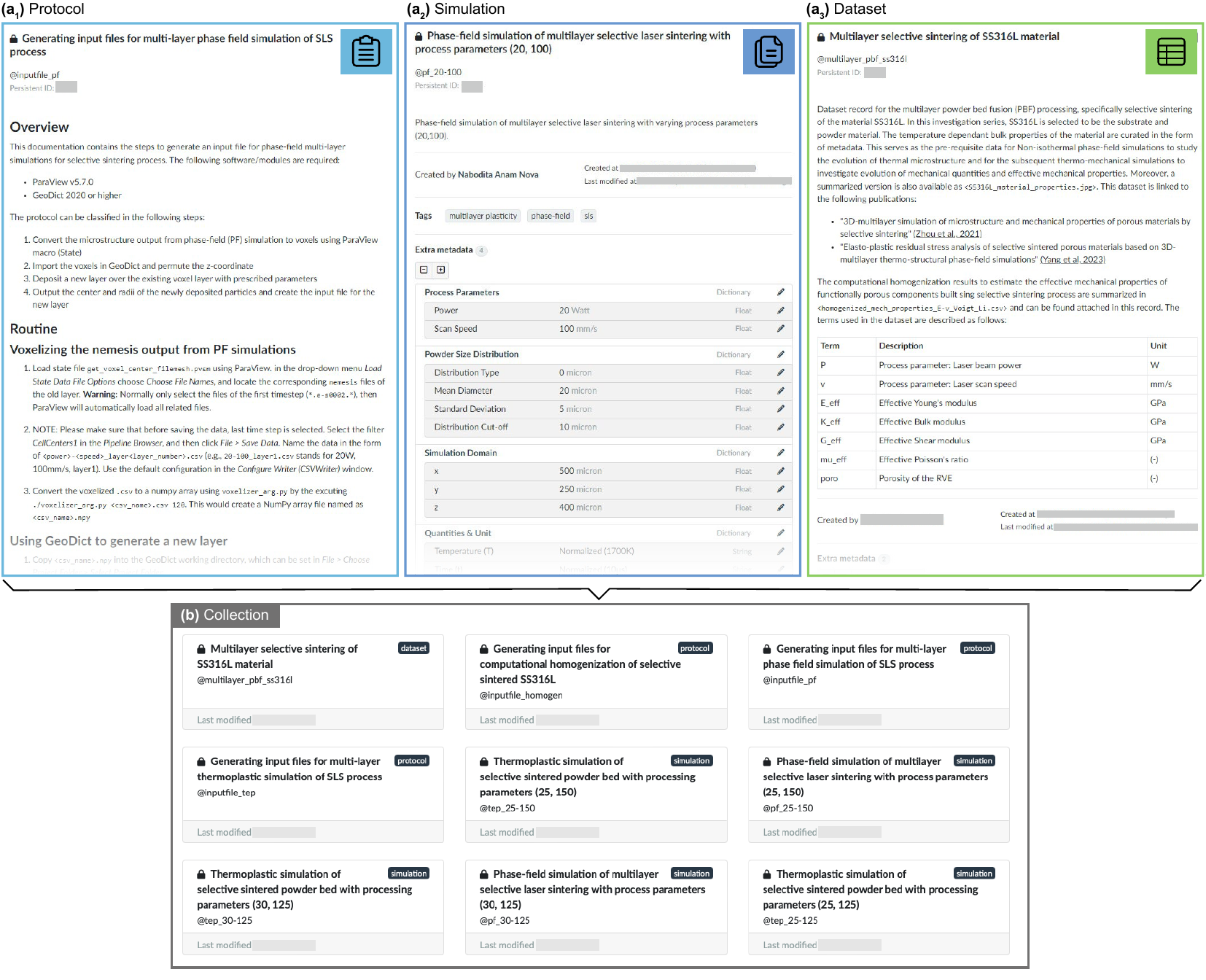}
    \caption{Snapshot of different record types: (a$_1$) Protocol record, (a$_2$) Simulation record and (a$_3$) Dataset record. These records along with other records form a collection as illustrated in (b).}
    \label{fig:records}
\end{figure}

The dataset records, for example \texttt{@mfm\_materials\_ss316l} contains the temperature-dependant material properties of the material SS316L that were used for phase-field simulations, thermo-mechanical simulations and computational homogenization, as shown in Fig. \ref{fig:records}(a$_3$). The material property data is stored in the metadata field. Protocol records, such as \texttt{@inputfile\_pf} utilize the description field to document the workflow to generate input files with a new layer of deposited powder particles, as shown in Fig. \ref{fig:records}(a$_1$). This particular workflow involves several sub-routines, including: conversion of phase-field based microstructure to voxel based microstructure, importing the voxelized microstructure into discrete element method (DEM) software (e.g., GeoDict\cite{geodict2023graingeo}, YADE\cite{yade:doc3}), depositing a new layer of powder over the previously processed layer, exporting the center and radii information of the newly formed powder bed and finally, using it to generate input file to process another layer of powder bed. The relevant metadata like software versions and powder characteristics are stored in the metadata field, while the supporting files like, macros and processing scripts are uploaded to the protocol record as attachments with their usage sufficiently documented in the description field. Fig. \ref{fig:records}(a$_2$) is an example of a simulation record corresponding to \texttt{@tep\_30-75} on the knowledge graph in Fig. \ref{fig:knowledge-graph}. Information like process parameters, powder bed characteristics, simulation domain size and the output quantities along with their normalized values and units are curated in the metadata fields, whereas the input files, containing the initial conditions, boundary conditions and the output result, in this case the localized temporal stress and strain in the microstructure is stored as file attachments within this record. The description field is employed to explain and/or document the usage of the files and the meta(data) values as well as the software versions used.

Records on Kadi4Mat have persistent and unique identifiers that distinguish them and ensure efficient retrieval and management. Upon creation, records are automatically assigned with a numeric PID, which remains unchanged. However, users can assign unique alphanumeric identifiers to further distinguish the records. The nomenclature for these identifiers is chosen to concisely represent the content of the record, such as material properties or simulation setup and results. In case of dataset records, the identifier name reflects the type of data as well as the material system. The identifiers of protocol records represent the procedure documented within them.
Data from the three types of simulations with varying process parameters is curated in form of simulation records. Their identifiers take the form \texttt{@<type>\_<P>-<v>}, to represent the simulation type as well as the distinguishing process parameters, in this case, the beam power \texttt{<P>} and scan speed \texttt{<v>}. \texttt{<type>} is a placeholder for the simulation type. For instance, \texttt{@pf\_20-100} would represent the simulation record for a multilayer phase-field simulation with beam power $20~\si{W}$ and scan speed $100~\si{mm~s^{-1}}$. Likewise, \texttt{@tep\_20-100} and \texttt{@homogen\_20-100} would be the records for thermo-mechanical simulation and computational homogenization respectively. 

Interlinking of records represents the relationships between them and the data flow within them as shown in the edges connecting the nodes in Fig. \ref{fig:knowledge-graph}. These linkages can be used to understand the overall simulation workflow. For instance, the protocol record \texttt{@powder-bed-gen} explains the procedure to generate the initial layer of powder bed. The centre and radii information of the initial powder bed is then exported to \texttt{@inputfile\_pf} for generating the input files to simulate the PBF process. \texttt{@inputfile\_pf} also receives temperature-dependant material properties of SS316L alloy from the dataset record \texttt{@mfm\_materials\_ss316l}. The input files generated from \texttt{@inputfile\_pf} is sent to the phase-field simulation records, which also store the simulated microstrutures of the processed layers. \texttt{@inputfile\_pf} adds another layer of powder on the previously processed microstructure and subsequently generates another input file and sends to the phase-field simulation records to process the PBF scan of new layer of powder. The transient thermal microstructure is delivered to the thermo-mechanical simulation records, where the relevant input files are sent by \texttt{@inputfile\_tep} upon recieving the material properties from \texttt{@mfm\_materials\_ss316l}. Likewise, for the computational homogenization, the final microstructure is transferred by the phase-field simulation records and the input file is created by \texttt{@inputfile\_homogen} to simulate the effective mechanical properties of the PBF processed parts with varying process parameters.

The research data curated from the aforementioned three simulation clusters is organised into collections on Kadi4Mat. Each collection is identified by the following identifiers: \texttt{@pf\_sls}, \texttt{@tep\_sls} and \texttt{@homogen\_sls}. These collections are published on Zenodo to boost the findability and accessibility of the curated research data.
The Zenodo entries contain records in JSON, RDF, and PDF formats, along with their corresponding metadata and file uploads. This comprehensive approach ensures that the records are stored in both human-readable and machine-readable formats, therefore, increasing their interoperability.
    
        
        
   
  
\subsection*{Automation of data recording integrated with high-throughput simulations}
Collecting and managing (meta)data from high-throughput investigations is of paramount importance for its further analysis and potential data-driven studies, however it can be a daunting task, if not automated. 
Automating the data collection step in a HTC workflow, not only boosts the efficiency of a laboratory, by eliminating the need for an intermediary (i.e. human operator), but also ensures consistency of the data records and compliance with the community-agreed standards. 

\begin{figure}
\centering
\includegraphics[width=\linewidth]{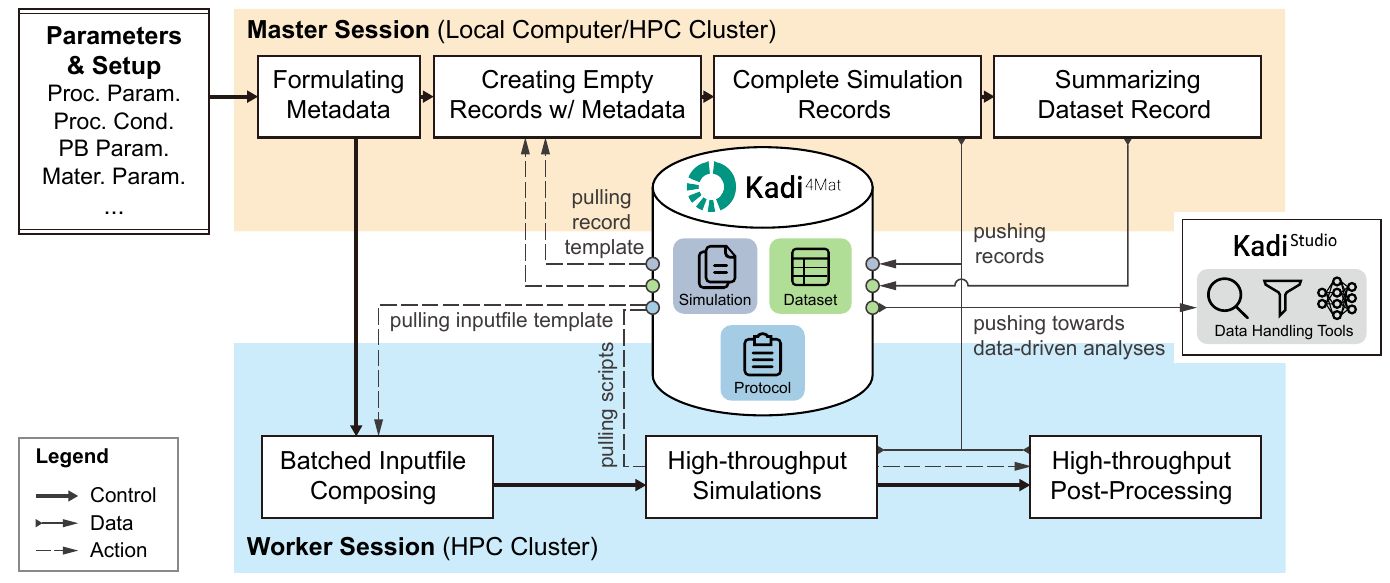}
\caption{Flowchart of integrated automation for performing high-throughtput simulation and post-processing together with RWDM framework.}
\label{fig:HTC}
\end{figure}

Fig. \ref{fig:HTC} depicts a typical HTC workflow, implemented on a master Jupyter session, with sub-routine scripts executed on the worker computer via ssh and the seamless interaction with Kadi4Mat powered by Kadi$^{\text{APY}}$. In a high-throughput investigation, batch simulations are performed with an array of combinations of processing parameters. The selection of process parameter combinations, often aided by design of experiments, is a critical step and depends on the processing window for the particular process and the process-material relationship. The process parameters and setup are normalized and formulated as metadata by the master console and are then fed into the inputfile composing scripts of the worker console along with the inputfile template pulled from the protocol records in data repository, to create batch inputfiles corresponding to the parameters. These batch inputfiles are submitted as batch jobs on the computing cluster, meanwhile, the master console pulls simulation record template from Kadi4Mat repository to create empty records for each simulation with the parameter information as metadata. Timed python scripts are used to check the status of the simulation jobs, completed jobs are verified for successful completion. The data from successfully completed jobs are further sent for processing, whereas unsuccessful jobs are resubmitted upon inspection. The post-processed data is pushed to their corresponding simulation records on Kadi4Mat. The data is further analyzed and summarized into a dataset, which would serve as an end-result of the HTC investigation. This dataset is also pushed as a dataset record in Kadi4Mat and could be retrieved for further data-cetric machine learning analysis. In this context, Kadi4Mat serves as a community repository as well as an ELN.
Alternatively, the data could be directly fetched into Kadi4Mat's ML workflow suite called KadiStudio\cite{griem2022kadistudio,tosato2023bayesian}.  KadiStudio has ML modules such as KadiAI and cids-tools to facilitate the development and implementation of data-driven models in ML workflows. These workflows can be documented as Kadi4Mat records to ensure their reproducibility.
  

\section*{Discussion}
The suggested RWDM framework was implemented for the simulation based multilayer PBF investigations. 
Standardization and consistency of the records was maintained by employing record templates. Python scripts based on Kadi$^\text{APY}$ library were used to create multiple records, making the framework capable of scaling up for even larger datasets, with the possibility of automating the data curation step, especially for HTC investigations. The current RWDM database for multilayer PBF studies can be expanded for an even wider range of processing parameters and can be extended to similar material systems. 

As the temporal thermal microstructure resulting from the process simulation was utilized as input in computational homogenization and thermo-mechanical analysis, they can be easily transferred for further scientific investigations such as nanoparticle migration behavior during PBF \cite{yang2021nanoparticle, gossling2023towards}, influence of process parameters on the magnetic properties of AM produced parts\cite{yang2023tailoring}, thermal anisotropy in porous AM parts\cite{yang2022diffuse}. Data from the Kadi4Mat records can be exported in machine-readable formats such as JSON and can be fetched automatically using Kadi$^\text{APY}$ based python scripts. The existing data can be extrapolated and utilised for further data-driven analysis as illustrated in Fig. \ref{fig:scheme}(b). These studies could possibly optimize the fused strut geometry by manipulating the volumetric energy input, predict the part properties for a particular set of processing parameters or tailoring the mechanical properties. Finally, reusing research data generated from computationally intensive simulations for further investigation is a step towards sustainable research.

The data curated through this RWDM workflow could later on serve as reference learning material for bachelor and master's degree students. Access to extensive research data can expedite the learning process for the students\cite{scheffler2022fair}. In conjunction with the learning process, access to a reference data would make it easier for other researchers to benchmark their simulation code and setup. The meticulously curated research workflow presented in this work underwent a rigorous test of reproducibility when a master's student was tasked with replicating the results solely by adhering to the documented procedures. With minimal guidance, the student successfully executed the workflows and reproduced the results, demonstrating the robustness and accessibility of the framework. The framework can be customized and extended to cater the data management needs of various other research works and could even facilitate collaboration within multiple levels of researchers. 
Principal investigators can initialize the customization of the RWDM framework by identifying the needs of their corresponding research projects. Followed by the breakdown of the project goals into tasks, and further breakdown into input, process and output components. These components can be managed using records, with detailed instruction of their usage along with automated macros/scripts in form of protocol records describing sub-workflows. Notably, it is important to set standards such as nomenclature of records, files and metadata descriptors, particularly in the case of collaborative research works. Upon customization of the framework, it can be implemented by research assistants/students or can be automated, especially for repetitive tasks, thereby accelerating the overall research workflow.
Implementing a similar RWDM framework would be a necessity for large projects like inter-laboratory study (ILS) involving numerous researchers from various research institutes, dealing with wide processing windows and multiple material systems\cite{kusoglu2021nanoparticle}.

\section*{Methods}
\subsection*{Data generation: Simulation scheme and workflow}
In this section, the simulation methods used to simulate the evolution of microstructure and the mechanical quantities (like stress and strain) during PBF processing are introduced. 
Based on our former research, a non-isothermal phase-field model was employed to simulate the microstructural evolution, while considering the aforementioned physical phenonema during single-scan multilayer PBF processing of SS316L parts \cite{yang2023elastoplastic, yang20193d, zhou20213d, yang2023tailoring}. 

The simulation scheme used to comprehensively investigate microstructural evolution and thermo-mechanical analysis is arranged in multiple stages, as shown in Fig. \ref{fig:scheme}(a). It starts with the parameterization and normalization of quantities used in the simulation models employed in this research work, such as phase-field parameters, temperature dependant material properties of SS316L and Argon atmosphere for thermo-mechanical analysis and the powder characteristics for powder bed deposition\cite{zhou20213d, yang2023elastoplastic}. 
A process window was selected with variation of the two most important process parameters for PBF: Beam power and scan speed. 
The non-isothermal simulator (NIsoS) program based on the MOOSE framework is employed to  implement the non-isothermal phase-field model using finite element method \cite{permann2020moose}. A simulation subdomain is selected and further imported into the thermo-elasto-plastic model and further into the computational homogenization scheme for calculating the mechanical quantities of the printed parts. The intricate details of the models used and the simulation workflow is sufficiently reported in our previous works\cite{yang20193d, zhou20213d, yang2023elastoplastic}.
\subsection*{Data sorting and organisation}
Diverse multifaceted data is generated throughout the simulation workflow as explained in previous section. The generated data is categorized in three main clusters, each representing a simulation stage from the workflow: (1) Phase-field simulations, (2) Thermo-mechanical simulations and (3) Computational homogenization.
Apart from the data generated from these simulations, there is a variety of supporting metadata with information crucial to reproduce these data. Therefore, it is imperative to identify, collect and organize the metadata from each stage of the simulations. Typical metadata for these simulations would be the material- and processing parameters, and their normalization, simulation setup such as, boundary conditions, initial conditions, numerical solver setup, powder characteristics for the generation of powder bed, software versions, etc. Furthermore, a comprehensive documentation of the sub-routines entailing the techniques and usage of software is necessary to achieve the reproducibility goal from the FAIR guiding principles\cite{wilkinson2016fair}.

\subsection*{Data recording and linking}
The huge variety of data generated from the series of simulations during the multilayer PBF investigation is recorded and curated in manner that it follows the FAIR principles\cite{wilkinson2016fair}. Kadi4Mat is employed to curate our research data in form of records. 
These records have various fields, such as title, identifier, description, metadata and file attachments to store the data and relevant metadata containing crucial information about the stored data. The records are uniquely identified using an alphanumeric identifier and a persistent numeric identifier. Standardization and consistency among the records are maintained by creating templates, which were then used to create records to store data from simulations with varying process parameters. Kadi4Mat offers the capability to store data in various record types, such as simulation records, protocol records and dataset records, as listed in table \ref{tab:record-types}. Protocol records were used to document the sub-routines and/or standard operating protocols (SOPs) used for a particular task from the workflow. Dataset records, as the name suggests, were used to store datasets, for example, material properties of SS316L used in the multilayer PBF investigation. The organised data from each simulation cluster was documented in form of simulation records using python scripts based on Kadi$^\text{APY}$. This ensured the scalability of our RDM framework by automating the data fetching and recording step to an extent. Linking of records with other records enables the visualization of data flow and exchange between the records, sometimes even depicting usage of particular data in the overall workflows.


\subsection*{Accessibility and publishing of data}
The records created on Kadi4Mat have the capability of being shared with the researchers within the Kadi4Mat consortia of institutions. However, for sharing the data with researchers outside the consortia, the records can be exported in various formats, including PDF, RDF and JSON. The records are also exported on Zenodo and is linked to a digital object identifier (DOI), thereby making the data findable and accessible to all. Zenodo supports data storage of up to 50 GB per record. For datasets exceeding this limit, multiple Zenodo records can be created or a third party data repositories can be utilized. 
\section*{Data availability}
The authors declare that the data curated using the RWDM framework described in this study for our multilayer powder bed fusion investigations are available on Kadi4Mat under the collection \texttt{@multilayer\_sls} with PID: 592.
An open access mirror copy of the summarized dataset is published on Zenodo and can be accessed via \url{doi.org/10.5281/zenodo.10940626}.

\section*{Code availability}

Source code for MOOSE-based application NIsoS and related utilities are available via the online repository \url{bitbucket.org/mfm_tuda/nisos.git}. The corresponding authors can be contacted for granting access. Exemplary python scripts, based on KadiAPY library, used to automate the data recording and fetching steps are available on Zenodo via \url{doi.org/10.5281/zenodo.8419354}.

\bibliography{sample}


\section*{Acknowledgements} 
Authors acknowledge the financial support of German Research Foundation (DFG) in the framework of the Collaborative Research Centre Transregio 270 (CRC-TRR 270, project number 405553726, sub-projects A06 and A07), and the Priority Program 2122 (SPP 2122, project number 493889809).
The authors also greatly appreciate the access to the Lichtenberg II high-performance computing (HPC) cluster and the technical support from the HHLR, Technische Universität Darmstadt. This work was partly carried out with the support of the Karlsruhe Nano Micro Facility (KNMFi, www.knmf.kit.edu), a Helmholtz Research Infrastructure at Karlsruhe Institute of Technology (KIT, www.kit.edu). The computing time on the HPC cluster is granted by the NHR4CES Resource Allocation Board under the project "special00007". S.B. also thanks the master's student Ms. Nabodita Anam Nova for her assistance in preparing showcase for proposed framework.

\section*{Author contributions statement}
Conceptualization: B.-X.X., Y.Y. and S.B.; methodology: S.B. and Y.Y.; software: S.B. and Y.Y.; data curation: S.B. and Y.Y.; visualization: S.B. and Y.Y.; writing-original draft preparation: S.B., Y.Y. and B.-X.X.; writing-review and editing: S.B., Y.Y., M.S., B.-X.X. and B.N.; supervision: B.-X.X.; consultation and discussion: M.S. and B.N.; funding acquisition: B.-X.X. All authors have read, reviewed and agreed to the published version of the manuscript.

\section*{Competing interests}
The authors declare no competing interests.


\begin{table}[ht]
\centering
\begin{tabular}{ll}
\hline
Record type & Data stored  \\
\hline
Protocol & Workflows, sub-routine, technical SOPs, relevant macros/scripts \\
Simulation & Input files, output files, relevant parameters  \\
Dataset & Material-specific parameters, data from secondary sources, aggregated output data\\
\hline
\end{tabular}
\caption{\label{tab:record-types}Record types and the corresponding research (meta)data stored.}
\end{table}

\end{document}